\documentclass{optica-article}

\journal{opticajournal} % for journals or Optica Open

\articletype{Research Article}

\usepackage{lineno}
\usepackage{graphicx}
\usepackage{subcaption}   % subfigure / subcaption
\usepackage{booktabs}

\usepackage{booktabs}     % Professional table rules
\usepackage{siunitx}      % Scientific formatting (S columns)
\usepackage{array}        % Advanced column types

\graphicspath{{figures0722/}}
\begin{document}

\title{GS-DOT: Gaussian splatting-based image reconstruction for diffuse optical tomography}

\author{Jingjing Jiang\authormark{1,*}}

\address{\authormark{1}Computational Optical Imaging Group, Biomedical Optics Research Laboratory, Dept. of Neonatology, University Hospital Zurich and University
of Zurich, Switzerland \\
%\authormark{2} 
}
\email{\authormark{*}jingjing.jiang@usz.ch} %% email address is required; see note below about the corresponding author designation

% use {asbstract*} to suppress the copyright line. Copyright information will be added in production

\begin{abstract*} 
This work presents GS-DOT, a novel image reconstruction framework based on Gaussian Splatting (GS) for diffuse optical tomography (DOT).  Inspired by GS for rendering applications, absorption coefficients are represented as a sparse sum of anisotropic Gaussian primitives optimized to fit measured time-resolved point-spread functions through analytic gradients and Adam optimization.  This is the first adaptation of GS algorithms in the photon diffusion regime, where the ray transport function is replaced by the diffusion equations to enable accurate modeling of light transport in highly scattering media. Validation on synthetic tissue models demonstrate high accuracy in localization and quantification of reconstructed absorption maps for both clean and noisy signals. GS-DOT has demonstrated high robustness to noise and showed a huge reduction in memory demand. 
\end{abstract*}

\section{Introduction}

Near infrared optical tomography (NIROT), or diffuse optical tomography (DOT), is a promising non-invasive imaging modality for providing valuable functional and structural information about biological tissues.
Image reconstruction for NIROT is an inverse problem, with the goal of recovering the optical property distributions $x \in X$ from boundary measurements $y \in Y$. The forward problem is formulated as $y = A(x) + \delta$, where $\delta$ is the noise in the measurement.  $A$ represents the forward model of light transport in tissues.  The traditional technique to solve this problem uses an iterative optimization-based algorithm \cite{Arridge09}. This process requires solving an inverse problem, which is non-linear, ill-posed and often under-determined. Model based methods are also computationally demanding and time-consuming \cite{dehghani2009near,Arridge09, Jiang20Aug}.

Recently, data driven approaches for image reconstruction, particularly deep learning (DL) have shown promising results in improving efficiency and image quality \cite{ben2018deep, yoo1990does,zou2021machine,benfenati2025modular,causin2019elastic,benfenati2020regularization, benfenati2013inexact}.
However, their performance often depends on large training datasets and lack of interpretability because many networks act as a black box and often ignore the underlying physics. These limitations reduce the reliability of the DL approaches for clinical applications. 

The recent gaussian splatting (GS), firstly introduced for efficient 3D scene representation in computer graphics \cite{kerbl23}. By representing volumetric space as discrete Gaussian elements, GS can increase the efficiency of light propagation modeling and enable implicit spatial regularization.
In the field of medical imaging, it has shown promising results in Xray CT \cite{cai2024}. 
Similar to GS for rendering applications \cite{kerbl23}, it also propagates a "signal" from a source through a medium to a sensor, weighted by a field function at each point in space.
 Unlike ballistic projection in rendering applications, diffusive propagation is used for diffusive media imaging. 
Using Gaussian Splats-representation naturally reduces dimension for the inverse problem solving, especially valuable for time-domain modalities \cite{Jiang20Aug, jiang2022resolution, Jiang20Oct}. Dimension reduction was accomplished previously by using a second coarser volume \cite{dehghani2009near} or anatomical priors  \cite{Barbour1995, Jiang17b,Jiang18B,Jiang18}.
The proposed GS based image reconstruction pipeline may also benefit other imaging modalities, for example, fluorescence imaging \cite{Ren20}, diffuse correlation tomography \cite{WANG2024}, electrical impedance tomography \cite{Mansouri2021}.

\section{Methods}

\subsection{Tissue representation using Gaussian splats}
% add a figure to illustrate 
Instead of the opacity and color representation used in rendering
applications \cite{kerbl23}, each Gaussian splat here carries a
single scalar parameter: the absorption contrast amplitude
$\alpha_k$  representing the peak absorption perturbation contributed by the $k-th$ basis function.
This substitution adapts the GS framework to the inverse 
problem in DOT, where the unknown is the spatially varying absorption
perturbation $\delta\mu_a(\mathbf{r})$ instead of a radiance field.
The $\delta\mu_a(\mathbf{r})$ can be formulated as superpositions of $K$ splats:
\begin{equation}
  \delta\mu_a(\mathbf{r}) = \sum_{k=1}^{K} \alpha_k
  \exp\!\left(-\frac{\|\mathbf{r} - \mathbf{c}_k\|^2}{2s_k^2}\right),
\label{eq:gs_iso}
\end{equation}
where $\alpha_k > 0$ is the peak amplitude, $\mathbf{c}_k \in \Omega$
is the geometrical center of the splat, and $s_k$ is the isotropic scale.
In the two-dimensional case, anisotropic GS with independently
adjustable axes and arbitrary orientation are used to allow the
reconstruction to adapt to elongated or irregularly shaped
inclusions. Each splat is evaluated in a locally rotated coordinate
frame $(u_k, v_k)$ aligned with its principal axes:
\begin{align}
 \delta\mu_a(\mathbf{r}) &= \alpha_k \exp\!\left(
    -\frac{1}{2}
    \left[
      \frac{u_k^2}{s_{x,k}^2}
      +
      \frac{v_k^2}{s_{y,k}^2}
    \right]
  \right),
\label{eq:gs_aniso}
\\[6pt]
  \begin{pmatrix} u_k \\ v_k \end{pmatrix}
  &= \mathbf{R}(\theta_k)\,(\mathbf{r} - \mathbf{c}_k),
  \qquad
  \mathbf{R}(\theta_k) =
  \begin{pmatrix}
     \cos\theta_k &  \sin\theta_k \\
    -\sin\theta_k &  \cos\theta_k
  \end{pmatrix},
\label{eq:gs_rot}
\end{align}
where $s_{x,k}$ and $s_{y,k}$ are the semi-axes along the rotated
$u$- and $v$-directions, and $\theta_k$ is the orientation angle.
All parameters are optimized in unconstrained form to facilitate
gradient-based optimization. The positivity constraints
$\alpha_k > 0$, $s_{x,k} > 0$, $s_{y,k} > 0$ are enforced
by implementing the following exponential form:
\begin{equation}
  \alpha_k = e^{\tilde{\alpha}_k}, \qquad
  s_{x,k}  = e^{\tilde{s}_{x,k}}, \qquad
  s_{y,k}  = e^{\tilde{s}_{y,k}},
\label{eq:gs_exp_params}
\end{equation}
while the central coordinates and orientation are optimized directly:
\begin{equation}
  \mathbf{c}_k = (x_k,\, y_k)^\top \in \mathbb{R}^2,
  \qquad
  \theta_k \in \left(-\tfrac{\pi}{2},\, \tfrac{\pi}{2}\right],
\label{eq:gs_direct_params}
\end{equation}
where $\tilde{\alpha}_k,\, \tilde{s}_{x,k},\, \tilde{s}_{y,k}
\in \mathbb{R}$ are real-valued variables.
The implementation of exponential re-parametrization guarantees strict positivity of the parameters at every gradient step.
The full parameter vector of all $6K$ unknowns is summarized as:
\begin{equation}
  \boldsymbol{\Theta} =
  \bigl\{
    \tilde{\alpha}_k,\; x_k,\; y_k,\;
    \tilde{s}_{x,k},\; \tilde{s}_{y,k},\; \theta_k
  \bigr\}_{k=1}^{K}
  \in \mathbb{R}^{6K}.
\label{eq:gs_theta}
\end{equation}

\subsection{Forward model}
 The forward problem of light-matter interaction was approximated by the diffusion equation (DE), which is valid when scattering dominates absorption ($\mu_s' \gg \mu_a$) \cite{Arridge99}. The re-emitted fluence rate $\Phi(\mathbf{r}, t)$ in the time-domain (TD) form at position $\mathbf{r}$ and time $t$ is described by:
\begin{equation}\label{eq:DE}
  [-\nabla\cdot D(\mathbf{r})\nabla+\mu_a(\mathbf{r})+\frac{1}{v_m}\frac{\partial}{\partial t}]\Phi(\mathbf{r},t)=q(\mathbf{r},t),
 \end{equation}  
where $v_m$ is the speed of light in the medium and $q$ is the isotropic source term. $v_m$ is the speed of light in the medium, $\mu_a$ is the absorption coefficient, $\mu_s'$  is the reduced scattering coefficient and $D$ is the diffusion coefficient  $D =[3(\mu_s')]^{-1}$ .
A GPU-facilitated finite element method (FEM)-based simulation tool NIRFASTer was used to solve the DE which is able to model complex geometries \cite{nirfaster,Doulgerakis17,Dehghani08}.

% add a figure to show differences between ray function (commonly used in volume rendering) and gaussian splatting

% - noise model
To assess the robustness of GS-DOT under realistic acquisition conditions, two scenarios are considered. In the noise-free case, the temporal point spread functions (TPSFs) are taken directly as the simulated forward model output $\Phi^\mathrm{clean}_{sd}(t)$.
 $1 \%$ Poisson  noise is applied to
$\Phi^\mathrm{clean}_{sd}(t)$ to generate noisy forward results $\Phi^\mathrm{noisy}_{sd}(t)$.

\subsubsection{Imaging geometry}
\begin{figure}[ht]
  \centering
  \includegraphics[width=0.85\textwidth]{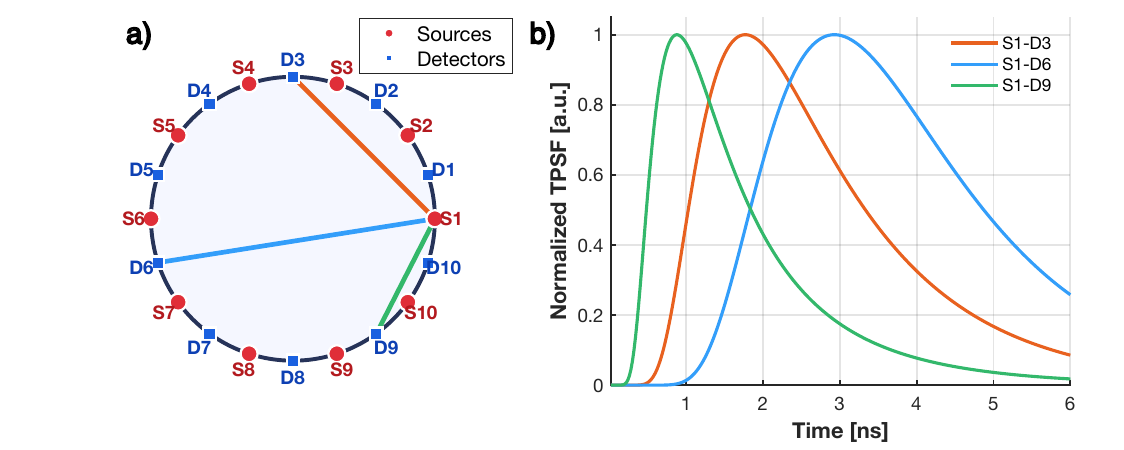}
  \caption{Measurement geometry. a) Ten source and ten detector optodes are
           distributed uniformly along the boundary of a circular domain of
           radius $R = 30$~mm, representative of a cross-section of a tissue volume. b) Temporal point spread functions (TPSFs) for three selected source-detector pairs. }
  \label{fig:geometry}
\end{figure}

The imaging geometry consists of a circular 2D domain of diameter $D = 6$~cm, which can represent cross-sections such as the neonatal head, with 10 sources and 10 detectors distributed evenly along the boundary (Figure \ref{fig:geometry}). Time-resolved measurements are acquired over a total temporal window of $T = 6$~ns, with temporal resolution of 
$\Delta t = 20$~ps. The temporal window is chosen to capture the full photon
arrival distribution including late arrival photons which carry the primary depth-sensitivity advantage of time-domain over continuous-wave DOT \cite{hebden1993time,Jiang20Aug,Jiang20Oct}. 
% Figure 1 — Measurement Geometry

\subsection{Inverse problem}
 \subsubsection{Sensitivity matrix}
The sensitivity or Jacobian matrix J reflects the changes in measured boundary fluence caused by small perturbation in tissue optical properties \cite{Arridge95}. Rytov approximation \cite{dehghani2009near} is used where the log-amplitude perturbation of the measured
TPSF $\Delta\Phi_{sd}(t)$ is linearly related to
the absorption perturbation $\Delta\mu_a(\mathbf{r})$:
\begin{equation}
  \Delta\Phi_{sd}(t)
  = \ln\left(\frac{\Phi^{\mathrm{meas}}{sd}(t)}{\Phi^{(0)}{sd}(t)}\right)
  = \int_\Omega J_{sd}(\mathbf{r}, t)\,
    \Delta\mu_a(\mathbf{r})\,\mathrm{d}\mathbf{r},
\label{eq:born}
\end{equation}
where $\Phi^{(0)}_{sd}(t)$ and $\Phi^{\mathrm{meas}}_{sd}(t)$ are the TPSFs for the baseline and target TPSFs,  and
$J_{sd}(\mathbf{r}, t)$ is the time-domain sensitivity matrix  is computed numerically using the finite-element method (FEM) solver implemented in NIRFASTer \cite{dehghani2009near}. For each source-detector (SD) pair $(s,d)$ and mesh node $\mathbf{r}$, the time-resolved sensitivity is obtained from the adjoint method as
\begin{equation}
J_{sd}(\mathbf{r}, t) = -\frac{v_m}{\Phi^{(0)}_{sd}(t)} \cdot
\bigl[\Phi_s(\mathbf{r}, \cdot)
* \Phi_d^{\dagger}(\mathbf{r}, \cdot)\bigr](t),
\label{eq:jacobian}
\end{equation}
where $*$ denotes temporal convolution, $\Phi_s(\mathbf{r}, t)$ is the time-domain photon fluence generated by source $s$ and propagated forward through the FEM-discretized mesh, and $\Phi_d^{\dagger}(\mathbf{r}, t)$ is the corresponding adjoint fluence generated by detector $d$ and propagated backward through the same mesh, both obtained by solving the time-dependent diffusion equation on the discretized domain using NIRFASTer's FEM framework. The full sensitivity matrix $\mathbf{J} \in \mathbb{R}^{(N_s N_d N_t) \times N_g}$
is precomputed once over all volume elements in $\Omega$.

\subsubsection{Optimization}
Before running the Gaussian splat optimizer, a fast backprojection image is computed by projecting the measured TPSF residual $\Delta\Phi_{sd}(t)$ onto the sensitivity
kernels across all SD pairs,
\begin{equation}
  \Delta\hat{\mu}_a^{\mathrm{BP}}(\mathbf{r})
  = \frac{\displaystyle\sum_{s,d,t}
          \Delta\Phi_{sd}(t)\cdot J_{sd}(\mathbf{r},t)}
         {\displaystyle\sum_{s,d,t}
          J_{sd}(\mathbf{r},t)^2 + \epsilon},
\label{eq:backprojection}
\end{equation}
where a small value $\epsilon$ was added to the denominator to avoid division by zero, which is likely to happen to those voxels outside the sensitive regions.
This gives a blurred but correctly localized first estimate of the perturbation positions. The centers of the splats were initialized using the K highest peaks through iterative peak finding. This provides a better initial parameter location for the GS-image reconstruction.  The unknown absorption perturbation is represented as a sparse sum of K anisotropic 2D GSs. The inverse problem in DOT is then solved by optimizing these parameters defining the GSs to fit the measured TPSFs. In this 2D case, we minimize the composite L2 loss across all 100 SD-pair TPSFs with the loss function defined as below.
\begin{align}
  \mathcal{L}(\boldsymbol{\Theta})
    &= \mathcal{L}_{\mathrm{data}}
     + \mathcal{L}_{\mathrm{reg}}
     + \mathcal{L}_{\mathrm{rep}},
  \label{eq:loss_total}
  \\[4pt]
  \mathcal{L}_{\mathrm{data}}
    &= \frac{1}{2\eta}
       \sum_{s=1}^{N_s}\sum_{d=1}^{N_d}
       \bigl\|\hat{\Phi}_{sd} - \Phi^{\mathrm{meas}}_{sd}\bigr\|^2,
  \qquad
       \eta = \max(\Phi^{\mathrm{meas}})^2 \cdot N_s N_d,
  \label{eq:loss_data}
  \\[4pt]
  \mathcal{L}_{\mathrm{reg}}
    &= \lambda_r \left[
         \sum_{k=1}^{K}\!\left(\tilde{\alpha}_k\right)^2
         +\, \beta
         \sum_{k=1}^{K}\!\left(\tilde{s}_{x,k} - \tilde{s}_{y,k}\right)^2
       \right],
  \label{eq:loss_reg}
  \\[4pt]
  \mathcal{L}_{\mathrm{rep}}
    &= \lambda_p
       \sum_{a=1}^{K}\sum_{b=a+1}^{K}
       \exp\!\left(
         -\frac{\|\mathbf{c}_a - \mathbf{c}_b\|^2}{2\rho_p^2}
       \right)
       \cdot
       \mathbf{1}\!\left[\|\mathbf{c}_a - \mathbf{c}_b\| < r_p\right],
  \label{eq:loss_rep}
\end{align}
with the full parameter vector defined in Eq. \ref{eq:gs_theta}. 
The term $\mathcal{L}_{\mathrm{data}}$ is the $\ell_2$ mismatch between the predicted TPSF $\hat{\Phi}_{sd}$ and the
target measurement $\Phi^\mathrm{meas}_{sd}$ summed over all 
$N_s \times N_d = 100$ source-detector pairs and $N_t$ time bins.
$\eta$ is a normalization factor that makes the loss dimensionless
and invariant with the signal amplitude and the size of the data set.
The regularization term $\mathcal{L}_{\mathrm{reg}}$ is added to smooth the optimization and it penalizes large log-amplitudes
$\tilde{\alpha}_k = \ln\alpha_k$ and excessive geometrical anisotropy through the log ratio of the two directions $(\tilde{s}_{x,k} - \tilde{s}_{y,k}) = \ln(s_{x,k}/s_{y,k})$.
 $\lambda_r$ and $\beta$ are the weights for  regularization and isotropy  penalty.
%acting as a log-Gaussian prior that encourages compact, approximately isotropic splats consistent with spherical tissue inclusions.
The repulsion term $\mathcal{L}_{\mathrm{rep}}$ penalizes spatial overlaps of nearby splats. The indicator function $\mathbf{1}[\cdot]$ is added to ensure the penalty is applied only to those pairs whose centers are separated by less than $r_p$. $\lambda_p$ is the repulsion weight and $\rho_p$ is the repulsion falloff scale. 

 % and $\rho_p = 0.4\,r_p$
 Adaptive moment estimation (Adam) optimizer with adaptive learning rates is used for updating the GS parameters \cite{kingma17ADAM}. After each iteration, the splat centers are  projected
onto the tissue domain $\Omega$ to enforce the physical
constraint $\mathbf{c}_k \in \Omega$. Optimizations are performed for both clean and noisy cases.

\subsection{Evaluation Metrics}

Reconstruction quality is evaluated quantitatively using three metrics: Root mean squared error (RMSE), Structural similarity index (SSIM), Center-of-mass localization error  $\varepsilon_\mathrm{CoM} $).
The RMSE measures the overall error of the reconstructed absorption map
compared to the ground truth across all  elements, and a lower value indicates lower error in the  reconstruction.
\begin{equation}
  \mathrm{RMSE} = \sqrt{\frac{1}{N_g}
  \sum_{i=1}^{N_g}\!\left(\hat{\mu}_a(\mathbf{r}_i)
  - \mu_a^*(\mathbf{r}_i)\right)^2}.
\label{eq:rmse}
\end{equation}
The SSIM assesses the quality of perceptual reconstruction by comparing luminance,
contrast, and structural similarity:
\begin{equation}
  \mathrm{SSIM}(X,Y) = \frac{(2\mu_X\mu_Y + c_1)(2\sigma_{XY} + c_2)}{(\mu_X^2 + \mu_Y^2 + c_1)(\sigma_X^2 + \sigma_Y^2 + c_2)},
  \label{eq:ssim}
\end{equation}
where $\mu$, $\sigma^2$, and $\sigma_{XY}$ are local means, variances, and
covariance, respectively, and $c_1$, $c_2$ are stabilization constants.
SSIM ranges from $-1$ to $1$, with $1$ indicating perfect reconstruction.

The localization accuracy of a reconstructed inclusion is assessed by
comparing the center of mass (CoM) of the reconstructed map to the
known inclusion center $\mathbf{c}^*$:
\begin{equation}
 \varepsilon_\mathrm{CoM} =
  \left\|\hat{\mathbf{c}} - \mathbf{c}^*\right\|_2.
  \qquad
  \hat{\mathbf{c}} =
  \frac{\displaystyle\sum_{i} \mathbf{r}_i\,\hat{\mu}_a(\mathbf{r}_i)}
       {\displaystyle\sum_{i} \hat{\mu}_a(\mathbf{r}_i)}, 
\label{eq:com}
\end{equation}
$\varepsilon_\mathrm{CoM}$ is reported in centimeters and reflects the
spatial displacement of the reconstructed inclusion from its true position.

\subsection{Case studies and image reconstruction results}
% ══════════════════════════════════════════════════════════════════════════════
% Figure — GS-DOT Reconstruction Results (4 cases × 2 columns)
% Column 1: Δμₐ maps (GT | Clean | Noisy)
% Column 2: Line profile through inclusion
% ══════════════════════════════════════════════════════════════════════════════
\begin{figure}[ht]
  \centering
  \includegraphics[width=1\textwidth]{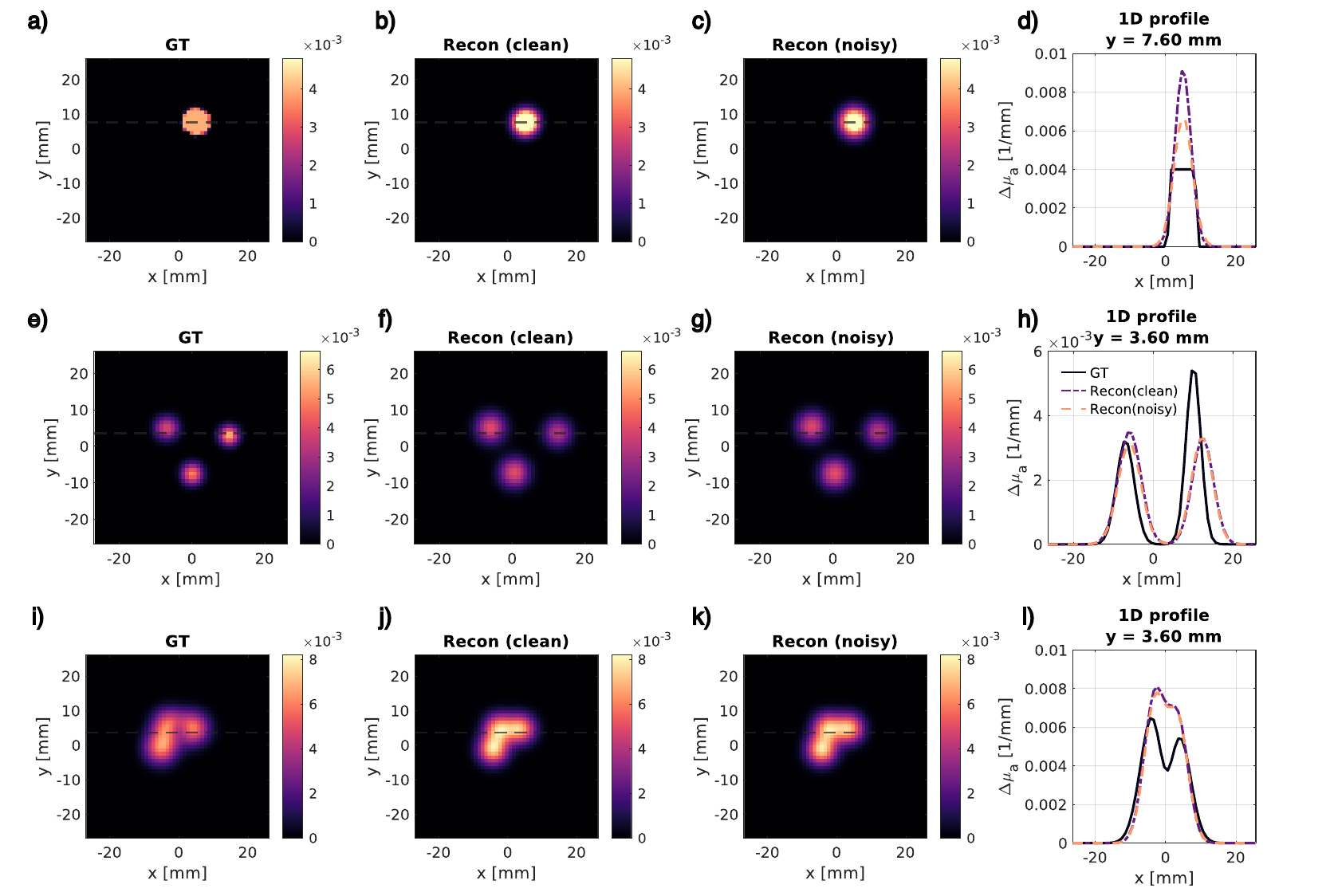}

  \caption{GS-DOT reconstruction results for three test cases.
           \textbf{Left column:} reconstructed $\Delta\mu_a$ maps showing
           ground truth (GT), clean reconstruction, and noisy reconstruction
           ($\eta_\mathrm{eff} \approx 1\%$).
           \textbf{Right column:}  $\Delta\mu_a$ line profiles through the
           inclusion centers comparing GT, reconstructions using clean  and
           noisy data.
           Rows correspond to (a-d) single circular inclusion,
           (e-h) three circular inclusions,
           (i-l) crescent-shaped inclusion.}
  \label{fig:results_rec_all}
\end{figure}
Three test cases were created to to evaluate the performance of GS-DOT, as illustrated in the left column of Fig. \ref{fig:results_rec_all}: 1) 1-inclusion 2) 3 inclusions 3) crescent shape.  The optical properties of the background medium are set to be $\mu_a = 0.001$~mm$^{-1}$, $\mu_s' = 1$~mm$^{-1}$, and refractive index $n = 1.4$ \cite{KHAN2021}.
A mesh of 5133 nodes was created to compute the TPSFs using Nirfast solvers.
The grid resolution is 1 mm. The active pixels within the circular domain is 2184,  and thus 2184 unknowns for conventional full grid-based image reconstruction. 

The reconstruction results for both clean and noisy signals  are shown in the second and third columns of Fig. \ref{fig:results_rec_all}.
The evaluation metrics are summarized in Table \ref{tab:reconstruction-metrics}. The metrics reveal little difference in the image reconstruction performance between  the noisy and clean signal cases.  Specifically, RMSE
variations remain low across all cases, SSIM values exceed $0.85$,and CoM localization errors are below $1$~mm.
This demonstrates the robustness of the GS-DOT. 
In terms of computational efficiency, GS-DOT reduces the parameter space from 2184 unknowns to 6K, where $K$ is in the range from $1$ and $6$ in the test cases, making the compression rates of 364 to 60.7. This enables efficient optimization while maintaining high fidelity in the reconstructed images.

%The computation was  entirely performed on CPU. 

\begin{table}[htbp]
\centering
\small
\caption{Quantitative evaluation metrics for GS-DOT reconstruction for the 3 test cases}
\label{tab:reconstruction-metrics}
\begin{tabular}{
  >{\raggedright\arraybackslash}p{1.8cm}
  S[table-format=1.3e-1]
  S[table-format=1.4]
  S[table-format=1.4]
  S[table-format=1.8,table-column-width=1.6cm]
  S[table-format=1.4]
  S[table-format=1.4]
}
\toprule
 & \multicolumn{3}{c}{\textbf{Clean}} & \multicolumn{3}{c}{\textbf{Noisy}} \\
\cmidrule(lr){2-4} \cmidrule(lr){5-7}
\textbf{Case} & {\textbf{RMSE}} & {\textbf{SSIM}} & {\textbf{$\varepsilon_\mathrm{CoM} $}} & {\textbf{RMSE}} & {\textbf{SSIM}} & {\textbf{$\varepsilon_\mathrm{CoM} $} } \\
 & {[mm$^{-1}$]} & & {[mm]} & {[mm$^{-1}$]} & & {[mm]} \\
\midrule
1 Inclusion  & \num{3.069e-4} & 0.966 & 0.116 & \num{2.358e-4} & 0.940 & 0.068 \\
3 Circles    & \num{3.265e-4} & 0.856 & 0.645 &\num{3.396e-4} & 0.855 & 0.644 \\
Crescent     & \num{3.648e-4} & 0.952 & 0.676 &  \num{3.570e-4} & 0.954 & 0.651 \\
\bottomrule
\end{tabular}
\end{table}

\section{Discussions}
All prior GS works assumes ballistic ray transport. Replacing the ray integral with the diffusion models is the core algorithmic novelty. 
One major motivation for adopting a GS-based parametric representation in DOT image reconstruction is the significant reduction in computational cost compared to conventional full-domain (voxel-grid). GS-based approach dramatically decreases the number of unknowns, therefore the memory requirements on the sensitivity matrix.
In the conventional voxel-grid approach, the unknown absorption perturbation $\Delta\mu_a(\mathbf{r})$ is discretized over all $N_g$ grid nodes inside the domain $\Omega$, yielding a parameter vector
$\boldsymbol{\mu} \in \mathbb{R}^{N_g}$. 
For example, for a $64 \times 64$ grid, $N_g = 4096$.
By contrast, the GS representation requires only $6K$ parameters for $K$. If $K = 6$, only 36 parameters are to be solved, which makes a compression ratio of approximately $114\times$ compared to the full grid methods. This ratio is expected to be even larger in three-dimensional cases.

%In terms of the storage of the sensitivity matrix $\mathbf{J} \in \mathbb{R}^{(N_s N_d N_t) \times N_g}$ can be precomputed and stored regardless of the reconstruction method. It requires memory of  $\mathcal{M}_\mathbf{J} = N_s \cdot N_d \cdot N_t \cdot N_g \cdot b_\text{float}$, where $b_\text{float} = 4$ bytes for single precision. For the case of ($N_s = N_d = 10$, $N_t = 300$, $N_g = 64^2$), it requires $\mathcal{M}_\mathbf{J} = 10 \times 10 \times 300 \times 4096 \times 4\;\mathrm{B} \approx 492\;\mathrm{MB}$.This precomputation process is a one-time cost. The critical difference lies in how this matrix is used during reconstruction.

In terms of computational cost at each iteration, in the full-grid methods, each gradient step requires evaluating the full matrix-vector product $\mathbf{J}\,\Delta\boldsymbol{\mu}_a$ and its transpose, with cost $\mathcal{O}(N_s N_d N_t \cdot N_g)$
per iteration. 
%Iterative solvers such as conjugate gradient (CG) or LSQR require $\mathcal{O}(N_\text{iter})$ such products.
%with $N_\text{iter}$ typically ranging from $10^2$ to $10^3$ \cite{Arridge91}.
In the GS-DOT approach, the updated forward result for each splat $k$ is the weighted product of the sensitivity matrix with
the splat profile evaluated at the pixels. The effective number of pixels is significantly lower than the total number of pixels $N_\text{eff} \ll N_g$ %nodes contribute significantly (within $\sim 3s_{x,k}$ of thecentre $\mathbf{c}_k$), 
giving an effective per-splat cost of $\mathcal{O}(N_s N_d N_t \cdot N_\text{eff})$.
%\begin{equation}
%  \hat{\Phi}_{sd}(t;\,\boldsymbol{\theta}_k)
%  = \int_\Omega J_{sd}(\mathbf{r},t)\,g_k(\mathbf{r})\,
%    \mathrm{d}\mathbf{r}
%  \approx \sum_{i=1}^{N_g}
 %   J_{sd}(\mathbf{r}_i, t)\,g_k(\mathbf{r}_i)\,\Delta A,
%\label{eq:gs_forward_cost}
%\end{equation}
%
The total forward updating cost is  $\mathcal{O}(K \cdot N_s N_d N_t \cdot N_\text{eff})$, which is substantially less than that using the full grid when $K \cdot N_\text{eff} \ll N_g$.

Beyond the computational efficiency, the GS parametrization also provides implicit regularization while full-grid methods often use Tikhonov regularization, L1, or total variation \cite{Dehghani08}.
Each Gaussian splat is inherently smooth, spatially compact, and non-negative, which aligns naturally with the optical models of biological tissues. This eliminates the need to tune
carefully regularization parameters to ensure the smoothness in the optimization to suppress artifacts, and in the same time , it does not sacrifice reconstruction fidelity.

One obvious limitation of this work is that the simple 2D tissue model is employed. Future works will use more complex 3D numerical models to tackle various imaging geometries \cite{Fang09, Dehghani08, Schweiger14}. This requires changes in the forward modeling and Jacobian calculation, while the framework remains the same. Secondly, the transmission-mode geometry was investigated in this study so that the initial estimation using Born back projection is valid. For reflection-mode imaging systems, different methods of initial estimation should be utilized. Another limitation is that only the absorption coefficient $\mu_a$ is investigated in this study, as it exhibits substantially greater spatial contrast than the reduced scattering coefficient $\mu_s'$ in near-infrared measurements \cite{Jiang20Aug}. Reconstruction of $\mu_s'$ remains for future work.

\section{Conclusion}
The Gaussian splatting-based image reconstruction algorithm GS-DOT has been developed for diffuse optical tomography. Unlike previous GS-approaches that rely on ray functions, diffusion models are employed for the highly scattering medium. It has been validated on synthetic data with and without noise and showed accurate reconstructed images in both cases. The GS representation significantly reduces parameter space, therefore, it requires less memory, imposes naturally the implicit regularization, and achieves higher efficiency in the inverse problem solving.
The quantitative evaluation showed its exceptional robustness against noise.

\begin{backmatter}

\bmsection{Acknowledgment}

 \label{sec:refs}
 \end{backmatter}
%%%%%%%%%% If using BibTeX:
\bibliography{sample}

\end{document}